\documentclass[a4paper]{article}

\usepackage[T1]{fontenc} 
\usepackage{float}       
\usepackage{helvet}      
\usepackage{mathptmx}    
\usepackage{rsc}         
\usepackage[sort&compress]{natbib}
\usepackage{setspace}    
\usepackage{graphicx}
\AtBeginDocument{\singlespacing}

\newfloat{chart}{htbp}{loc}
\newfloat{graph}{htbp}{loh}
\newfloat{scheme}{htbp}{los}

\usepackage[version=3]{mhchem} 

\long\def\symbolfootnote[#1]#2{\begingroup%
\def\thefootnote{\fnsymbol{footnote}}\footnote[#1]{#2}\endgroup} 

\author{%
  Dario Marrocchelli \thanks{%
    School of Chemistry, The university of Edinburgh, West Mains Road EH9 3JJ, UK
    E-mail: \texttt{D.Marrocchelli@gmail.com }}%
  \and
  Paul A Madden\thanks{%
    Department of Materials, University of Oxford, Parks Road, Oxford OX1 3PH, UK,
  }%
  \and
  Stefan T Norberg\thanks{%
    Department of Chemical and Biological Engineering, Chalmers University of Technology, SE-412 96 Gothenburg, Sweden
  }%
 \and
 Stephen Hull \thanks{%
 The ISIS Facility, Rutherford Appleton Laboratory, Chilton, Didcot, Oxfordshire OX11 0QX, UK}
}

\title{Vacancy ordering effects on the conductivity of yttria- and scandia-doped zirconia.}

\begin{document}

\maketitle

\begin{abstract}
Polarizable interaction potentials, parametrized using ab initio electronic structure calculations,
have been used in molecular dynamics simulations to study the conduction mechanism in Y$_2$O$_3$- and Sc$_2$O$_3$-doped
zirconias. The influence of vacancy-vacancy and vacancy-cation interactions on the conductivity of these materials has been characterised.
While the latter can be avoided by using dopant cations with radii which match those of Zr$^{4+}$ (as is the case of Sc$^{3+}$), the former
is an intrinsic characteristic of the fluorite lattice which cannot be avoided and which is shown to be responsible for the occurrence of a maximum in the conductivity at dopant concentrations between 8 and 13 \%. The weakness of the Sc-vacancy interactions in Sc$_2$O$_3$-doped zirconia suggests that this material is likely to present the highest conductivity achievable in zirconias.
\end{abstract}

\section{Introduction}

Yttria-doped zirconia (YSZ) is a fluorite-structured oxide ion conductor used as an electrolyte in a solid-oxide fuel cell (SOFC) {\it inter alia}. Substitution on the cation sublattice of the Zr$^{4+}$ions by the larger Y$^{3+}$ ions stabilizes the eight-coordinate cubic structure and, because of the lower valency, introduces oxide ion vacancies on the anion lattice which then allow oxide ion diffusion \cite{hull2004}. If diffusion occurred by an independent vacancy-hopping mechanism, one might expect that the dependence of the conductivity on vacancy concentration, $c$, would be $c(1-c)$, which maximizes at $c=0.5$. However, the maximum actually occurs at a much lower vacancy concentration -- where roughly 4\% of the oxide ion lattice sites are unoccupied \symbolfootnote[5]{If the compositions is denoted (M$_2$O$_3$)$_{(1-x)}$(ZrO$_2$)$_x$, where M is a trivalent cation, then the fraction of unoccupied anion lattice sites, or vacancy concentration, is given by ${\frac {x}{2(1+x)}}$}. This has been ascribed to the reduction in the vacancy mobility by ordering processes. These can be caused by the differences between the host and dopant cations, which lead to a preference for vacancies to bind close to a certain cation species \cite{arachi1999, khan1998, zacate1999}, or by the interaction between the vacancies themselves \cite{goff1999, bogicevic2003}. \newline

In the case of YSZ, adding Y$^{3+}$ ions has two competing effects: on the one hand, adding more dopant increases the number of vacancies and therefore increases the mobility of the remaining anions; on the other hand, since the migration barrier of an oxygen crossing a Y-Zr or Y-Y edge is much higher than in the case of a Zr-Zr edge \cite{kilo2003, krishnamurthy2004, kushima2009} (because Y$^{3+}$ is a much bigger cation \cite{shannon1976} than Zr$^{4+}$), adding more dopant will increase the fraction of cation edges associated with high migration energy barriers, thus effectively reducing the overall number of {\em mobile} oxygens. It is worth noting that this can be seen from a vacancy perspective: vacancies prefer to bind to be nearest neighbours to the smaller cations (Zr$^{4+}$ in this case) and this reduces their overall mobility. If we express the oxygen diffusion coefficient, $D$, as \cite{devanathan2006}
\begin{equation}
D=D_0\;exp{(-\frac{\Delta H_m}{k_BT})}
\end{equation}
where $D_0$ is a pre-exponential factor, $T$ the temperature in Kelvin, $k_B$ Boltzmann's constant and $\Delta H_m$ is the oxygen migration enthalpy, increasing the number of vacancies will increase $D_0$ whereas increasing the number of Y-Zr or Y-Y edges will increase the migration enthalpy, $\Delta H_m$. The resulting diffusion coefficient will peak for a certain dopant concentration, which is smaller than would be expected from an independent vacancy-hopping mechanism. This explanation is almost universally accepted \cite{sawaguchi2000, kilo2003, krishnamurthy2004, pornprasertsuk2005, devanathan2006, martin2006, devanathan2009} and has been generalised to other doped fluorite systems \cite{krishnamurthy2004, martin2006, devanathan2009, sato2009}. In  scandia-stabilised zirconia, however, the two cations have very similar radii and cation-vacancy ordering effects are very weak \cite{khan1998, zacate1999, devanathan2009, norberg2010}. For this reason the drop in the conductivity observed in this material (at a slightly greater dopant concentration than in YSZ) has to be caused by some other effect.
 \newline

Vacancy-vacancy interactions have also been proposed as a factor hindering the ionic conductivity of these materials \cite{goff1999, irvine2000, garcia2000, garcia2005}. Direct evidence of vacancy-vacancy ordering even at the high temperatures of interest for conductivity studies, is seen in diffuse neutron scattering patterns obtained for YSZ \cite{goff1999}. In this study, intense diffuse peaks are observed in the neutron and x-ray diffraction patterns at positions which are forbidden for a simple fluorite lattice. Some of these peaks have been interpreted as caused by the local relaxation around small aggregates in which pairs of vacancies along the $\langle111\rangle$ direction pack together along the $ \langle112\rangle$ direction. That this vacancy pair configuration is favourable is validated by the {\it ab initio} static energy calculations of Bogicevic and co-workers \cite{bogicevic2003} and Pietrucci {\it et al.} \cite{pietrucci2008}. By studying the relaxation times observed in the quasielastic neutron scattering at different points in reciprocal space, Goff {\it et al.} showed that the vacancies associated with the aggregates moved more slowly than isolated vacancies and suggested that this was the cause of the decrease in the conductivity at high vacancy concentrations. \newline

In the preceding paper we examined cation-vacancy and vacancy-vacancy ordering effects in the Zr$_{0.8}$(Y/Sc)$_{0.2}$O$_{1.9}$ system, using computer simulations. We studied systems in which the proportion of Y$^{3+}$ and Sc$^{3+}$ is varied at {\em constant} vacancy concentration (this is 5\%, close to the composition at which the conductivity maximum occurs in YSZ). The simulations reproduced the structural information from diffraction experiment and also the variation of the conductivity as the Y/Sc ratio was varied. We found evidence of both cation-vacancy and vacancy-vacancy ordering and the associated energies were consistent with the ab-initio data \cite{bogicevic2003}. The cation-vacancy effects were much stronger in the Y-rich samples, which is consistent with the lower conductivity of those materials, whilst the vacancy-vacancy ordering seemed to be almost independent of the nature of the dopant cation, at least at these dopant levels. \newline


The objective of the present paper is to examine the interplay between these two mechanisms and to discover how they combine to determine the conductivity of a particular material at a particular doping level. We will examine the properties of (Y$_2$O$_3$)$_x$ - (ZrO$_2$)$_{1-x}$ and (Sc$_2$O$_3$)$_x$ - (ZrO$_2$)$_{1-x}$ (which we dub ScSZ) obtained in simulations at high temperature, with $x$ varying and, with it, the vacancy concentration. The ultimate purpose is to elucidate to what extent it might be possible to further improve the conductivities of this class of  material by further optimisation of the doping strategy.  Although both mechanisms clearly exist, it is not clear to what extent they are coupled. Does the cation-vacancy interaction influence the vacancy-vacancy interaction to a significant degree, for example? Our strategy is, firstly, to compare the results of our simulation with measurable quantities which bear upon the properties of the vacancies, notably the conductivity and diffuse scattering. No new experimental data is reported in the present paper, the conductivities of both systems have been measured previously \cite{kilner1982, arachi1999} and the diffuse neutron scattering in the YSZ system was measured in {\it single crystal} studies by Goff {\it et al.} \cite{goff1999} for $0.09<x<0.25$. Secondly, having validated that our simulations are reproducing these observables sufficiently closely to assert that the behaviour of our simulated systems is paralleling that of the real materials in this regard, we will examine directly the properties of the vacancies themselves in a way which cannot be replicated in a real experiment. In addition to the simulation of these materials with realistic potentials, we will also report results for model systems for which the potentials have been altered (for example, by equalization of cation charges) in order to illustrate the physical factors responsible for a particular aspect of the observed behaviour.  \newline

 \section{MD simulation and analysis}
The interaction potential (which we usually dub DIPPIM), and the procedure we used to parameterize it, has already been described in preceding work \cite{norberg2009a, marrocchelli2009a, marrocchelli2009b}. In this potential, the ionic species carry their valence charges (Zr$^{4+}$, Y$^{3+}$, Sc$^{3+}$ and O$^{2-}$), and the polarization effects that result from the induction of dipoles on the ions are accounted for. The parameters of the interaction potential for these systems were obtained from the application of a force-and dipole-matching procedure aimed at reproducing a large set of first-principles (DFT) reference data \cite{madden2006a} on the condensed phase. Such potentials have been shown to provide an accurate and transferable representation of the interactions in a number of oxides \cite{wilson2004, madden2006a, jahn2007b,norberg2009a, marrocchelli2009a, marrocchelli2009b, marrocchelli2009c, marrocchelli2010a}. \newline

In addition to these ``realistic" potentials we also construct "ideal" model systems (which we will refer to as i-YSZ$_x$) which contains the same concentration of vacancies as (Y$_2$O$_3$)$_x$ - (ZrO$_2$)$_{1-x}$ with the same $x$ value. In i-YSZ$_x$  {\em all} the cations carry the same charge, such that the total cation charge balances that of the O$^{2-}$ ions present in the simulation, and all cations have the same short-range interaction potentials as the Zr$^{4+}$ ions in the original potential. This idealized system allows us to eliminate the effects of differing charges and lattice strain induced by having host and dopant cations and gives an idealised reference system in which the cation-vacancy ordering effects are absent. \newline

All the simulations were performed, unless otherwise stated, using a cubic simulation box with
4 x 4 x 4 unit cells of the fluorite structure, i.e. 256 cations and a variable number of oxygen ions, depending on cation composition. For each dopant concentration, a certain number of Zr$^{4+}$ ions were replaced with Sc$^{3+}$ or Y$^{3+}$ ions and their positions were randomly distributed over the cation sublattice \cite{norberg2009a, marrocchelli2009b}. In our previous work \cite{marrocchelli2009b} we concluded that local cation ordering, as observed in Y$_3$NbO$_7$, was much less important in Y$_2$Zr$_2$O$_7$ and we presume that this remains true at other compositions as well as in ScSZ (where the site mismatch between Sc$^{3+}$ and Zr$^{4+}$ is even smaller than in the Y$^{3+}$ and Zr$^{4+}$ case \cite{shannon1976} ), so that a random distribution of cations is appropriate. Vacancies were randomly assigned to the oxide sublattice in the initial configuration. The time step used was 1 fs and all the runs were performed at constant volume and temperature (NVT ensemble), with thermostats as described elsewhere \cite{martyna1994a}. The cell volume was obtained from a previous run in a NPT ensemble with zero applied pressure. Coulombic and dispersion interactions were summed using Ewald summations while the short-range part of the potential was truncated to half the length of the simulated box (usually about 10 \AA).\newline

In order to calculate ionic conductivities, we ran long simulations (0.5 $\div$ 5 ns) at high temperatures (T $>$ 1250 K) on YSZ and ScSZ. Each simulation was long enough to allow each oxygen ion to move, on average, by a distance of at least $\sim 2.6$ \AA, i.e. the oxygen-oxygen bond distance. Ionic conductivities were  estimated from the slope of a plot of the mean-square displacement of the {\em charge} {\it versus} time, {\it i.e.} \cite{castiglione2001}
 \begin{equation}
 \lambda = \frac{1}{6k_BTV}\; \lim_{t \to \infty}  t^{-1} \langle \mid \sum_i q_i\delta {\bf r}_i(t) \mid ^2\rangle, \label{real-conductivity}
 \end{equation}
 where $q_i$ is the charge on ion $i$, $\delta {\bf r}_i(t)$ the displacement made by the ion in time $t$ and $\langle$....$\rangle$ denotes an average over the simulation run. This quantity suffers from poor statistics, especially when the mobility of the ions is low, and so we have also calculated conductivities from the Nernst-Einstein expression
 \begin{equation}
 \lambda = \frac{1}{6k_BTV}\; \lim_{t \to \infty}  t^{-1}   \sum_i q_i^2\langle \mid \delta {\bf r}_i(t) \mid ^2\rangle,
 \end{equation}
 which neglects correlations between the diffusive jumps of different ions. \newline

To compare with the diffuse neutron diffraction results we calculate the intensity of
(total) scattering at some point ${\bf q}$ in reciprocal space from
\begin{equation}
I({\bf q})=\langle \sum_i \sum_j a_ia_j^{\star} e^{i{\bf q}\cdot{\bf r}_{ij}}
\rangle; \label{diffuse}
\end{equation}
where the sums run  over all ions in the sample, ${\bf r}_{ij}$ is the vector joining ions $i$ and $j$ and $a_i$ is the neutron scattering length of the species to which $i$ belongs. Because of the periodic boundary conditions, the accessible vectors ${\bf q}$ with a cubic simulation cell of side $L$ are restricted to the set ${\frac {2\pi}{L}}(m,n,p)$ where $m$, $n$, and $p$ are integers, and this provides a limit to the resolution of the simulated pattern. For these calculations we used a cubic simulation box with 6 x 6 x 6 fluorite unit cells, which gives a resolution of approximately 0.2 \AA$^{-1}$. We made extensive use of the cubic symmetry to average over the scattering calculated from symmetry-related ${\bf q}$-vectors. These simulations were started at 1800 K and then slowly cooled down to 800 K with a cooling rate of 10$^{12}$ K s$^{-1}$. In this temperature range, oxygen ion diffusion is observed while, below 800 K, no diffusion can be observed on the available timescale. For this reason the simulations at low dopant content ($x$ = 0.9) were further quenched down to room temperature (300 K). Although this should not influence the vacancy ordering this will increase the tetragonal distortions in the sample (see discussion below).\newline

The relaxation of the structures responsible for the diffuse scattering at a particular point in reciprocal space ${\bf q}$ can be measured in a quasielastic neutron scattering experiment. In the simulation the quantity to be calculated for comparison with the experimental data is the intermediate scattering function $I({\bf q},t)$, given by
\begin{equation}
              I({\bf q},t)=\langle \sum_i \sum_j a_ia_j^{\star} e^{i{\bf q}\cdot{\bf r}_{i}(t)}
e^{-i{\bf q}\cdot{\bf r}_{j}(0)}\rangle,
\end{equation}
again, the calculated quantities may be averaged over the symmetry-related $\bf{q}$-vectors.
\newline

We can identify the positions of vacancies for some instantaneous ionic configuration by finding which of the coordination tetrahedra around the anion sites are empty. The details of this procedure are reported in previous papers \cite{castiglione2001, marrocchelli2009b}. Because the cations are not diffusing we can monitor the properties of each tetrahedron from the identities and instantaneous positions of the four cations which sit at its vertices. Such a tetrahedron is empty if no anion is within the volume bounded by the four planes defined by the positions of each set of three of the cations. Because the oxide ions may perform large-amplitude vibrations about their average sites at the temperatures of interest for dynamical studies we only assign a vacancy to a tetrahedral site if the tetrahedron has been empty for a  minimum of two frames (i.e.\ 100 fs). The position of the vacancy is defined by that of centre of the tetrahedron (which is given by the average positions of the four surrounding cations). Once the positions of vacancies are identified, we can us them to build radial distribution functions (rdfs) to study the ordering tendencies in real space. Integration of the rdf can be used to define coordination numbers. For example integrating  the vacancy--vacancy rdf, g$_{V-V}$ from zero out to the position $r_c$ of first minimum of the g$_{V-V}$ rdf gives the average number of vacancies which are nearest-neighbours another vacancy:
\begin{equation}
n_{V-V}= 4\pi \rho \int_{0}^{r_c}dr\; r^2\; g_{V-V}(r),
\end{equation}
where $\rho$ is the density of vacancies.

\section{Comparison with experimental data}
In this section we will compare our MD data with the available experimental data on (Y$_2$O$_3$)$_x$-(ZrO$_2$)$_{1-x}$ and (Sc$_2$O$_3$)$_x$-(ZrO$_2$)$_{1-x}$. Although our methodology has already been thoroughly tested in the preceding paper \cite{norberg2010} as well as in our previous work \cite{norberg2009a, marrocchelli2009b} for $x$ = 0.11, $x$ = 0.33, we will first ensure that our simulations at different dopant concentrations reproduce the experimental data which is influenced by the vacancy ordering effects, before drawing conclusions on the conduction mechanism of YSZ and ScSZ. A comparison will be therefore made with the conductivity, quasielastic and diffuse scattering data \cite{goff1999, kilner1982, subbarao1979, nakamura1986}.

 \subsection{Conductivity}

\begin{figure}[htbp]
\begin{center}
\includegraphics[width=10cm]{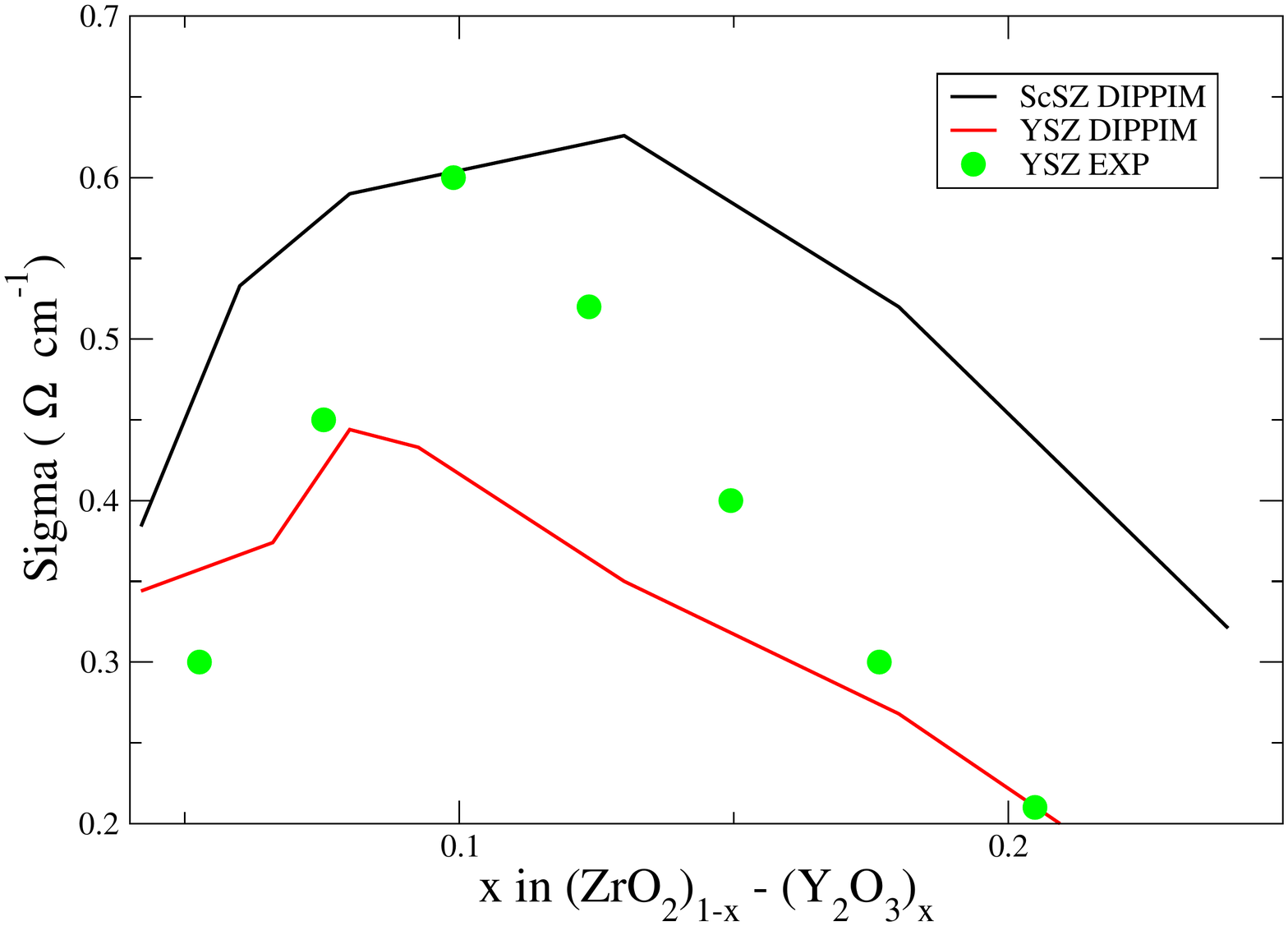}
\includegraphics[width=10cm]{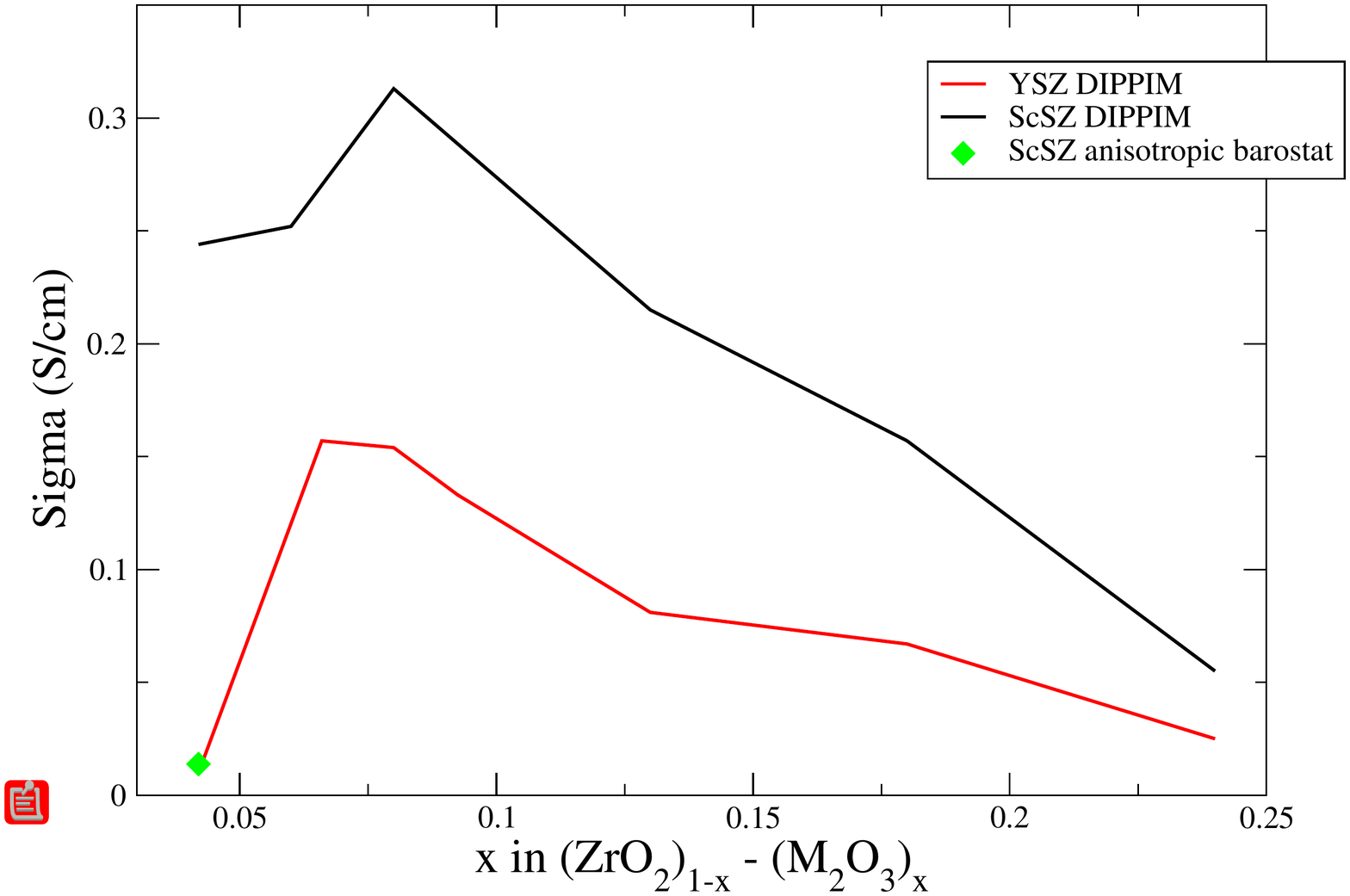}
\end{center}
\caption[Conductivity vs dopant concentration for YSZ and ScSZ at 1670 K.]{\it{\small Top: MD conductivities vs doping (x) for YSZ (red curve) and ScSZ (black curve) at 1670 K. We also report the experimental data from \cite{subbarao1979, nakamura1986} for YSZ at the same temperature. Bottom: MD conductivities vs doping (x) for YSZ (red curve) and ScSZ (black curve) at 1250 K.}}
\label{SIGMA_MD}
\end{figure}

In figure \ref{SIGMA_MD} we show the calculated Nerrnst-Einstein conductivities for both YSZ and ScSZ as a function of dopant concentration at 1670 K. For YSZ, we also report the experimental values from ref. \cite{subbarao1979, nakamura1986}. The MD data on YSZ are seen to be in reasonable quantitative agreement with experiment. The dependence of the conductivity on the dopant concentration also looks promising. Both curves show a peak in the conductivity at $x$ = 8-9 \% and 12-13 \% for YSZ and ScSZ respectively in good accord with the experimental data. However, the rise in conductivity shown by our simulations is smaller than that of the experimental data. This may  be caused by two effects. Firstly, at higher dopant concentrations, the Nernst-Einstein approximation of independent ion jumps is not accurate. To this end, we calculated the ionic conductivities from equation \ref{real-conductivity} at the dopant concentrations shown in figure \ref{SIGMA_MD}, and found that these are in close agreement with the Nernst-Einstein values for $x<$ 0.10 but systematically lower to an increasing degree as $x$ increases, for $x>$0.10. Secondly, the small systems studied in this paper, which do not allow the formation of grain boundaries and tetragonal domains, might affect the behaviour of the low dopant concentration part of the curve  (see also discussion below).
\newline

The conductivities shown above were calculated from simulations in which the cubic fluorite symmetry was enforced. This was done by equilibrating the simulation using an isotropic barostat \cite{martyna1994a}, before conducting the production runs at constant volume. As it is well known \cite{goff1999, hull2004}, YSZ and ScSZ are not cubic at low temperatures and dopant concentrations. For this reason, when the conductivity versus dopant concentration curves are calculated at lower temperatures, a slightly different behaviour is observed (see bottom part of figure \ref{SIGMA_MD}). The maximum has, in fact, shifted to lower dopant concentrations (approximately 6 \% and 8 \% for YSZ and ScSZ respectively) and the curve's shape looks very similar to that of Gd$_2$O$_3$-doped ZrO$_2$ (see figure 4 in reference \cite{arachi1999}, a material which, due to Gd's big ionic radius, is cubic for almost all compositions). A shift of the maximum's position as the temperature is lowered is also observed experimentally \cite{subbarao1979, nakamura1986}, though not as strong as that indicated by the comparison of the top and bottom part of figure \ref{SIGMA_MD}. \newline

We think that the changes observed in figure \ref{SIGMA_MD} are mainly caused by the fact that the real material is not locally cubic at these temperatures and low dopant concentrations\cite{goff1999, hull2004} , so that the simulations, in which the cubic symmetry is enforced within a cell which is too small to allow the formation of distorted domains, tend to overestimate the conductivity of these materials. To confirm this, we relaxed the cubic fluorite symmetry in a simulation of 4\% ScSZ and allowed the simulation cell to change its shape, by using an anisotropic barostat \cite{martyna1994a}. The simulation quickly adopted a non-cubic symmetry. The conductivity obtained for this system is shown in the bottom part of figure \ref{SIGMA_MD} as a green diamond. It can be appreciated that this is indeed much lower (more than ten times!) than the conductivity obtained from a cubic simulation. In the reminder of this paper, however, we will focus on highly-doped stabilised zirconias only, i.e. those materials with more than 8 \% of dopant, which are fully stabilised at every temperature. These systems are therefore not affected by the above-mentioned problem and therefore simulations, in which the cubic symmetry is enforced, can be used.
\newline

 \subsection{Diffuse scattering in YSZ}

\begin{figure}[htbp]
\begin{center}
\includegraphics[width=6.0cm]{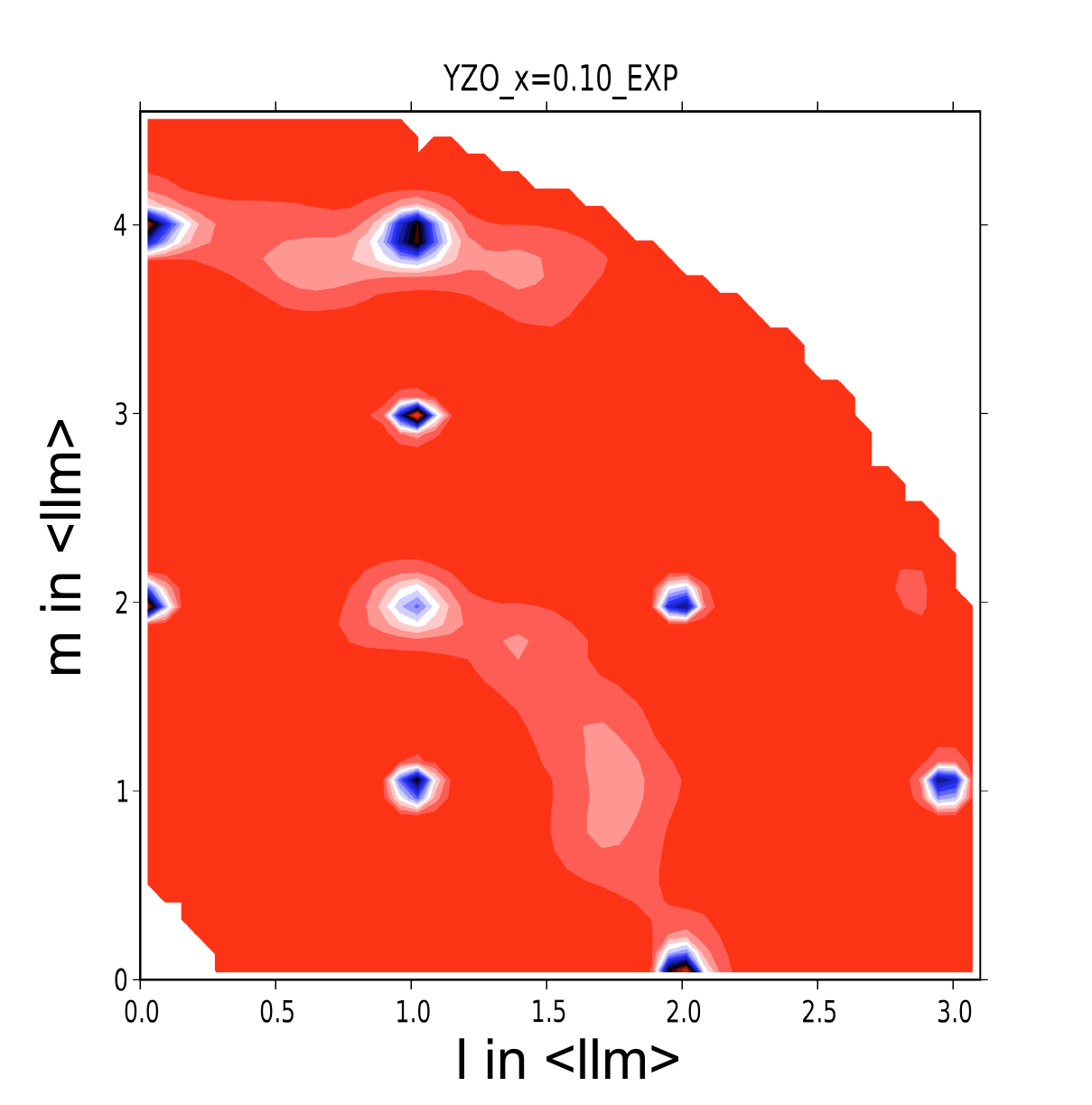}
\includegraphics[width=6.0cm]{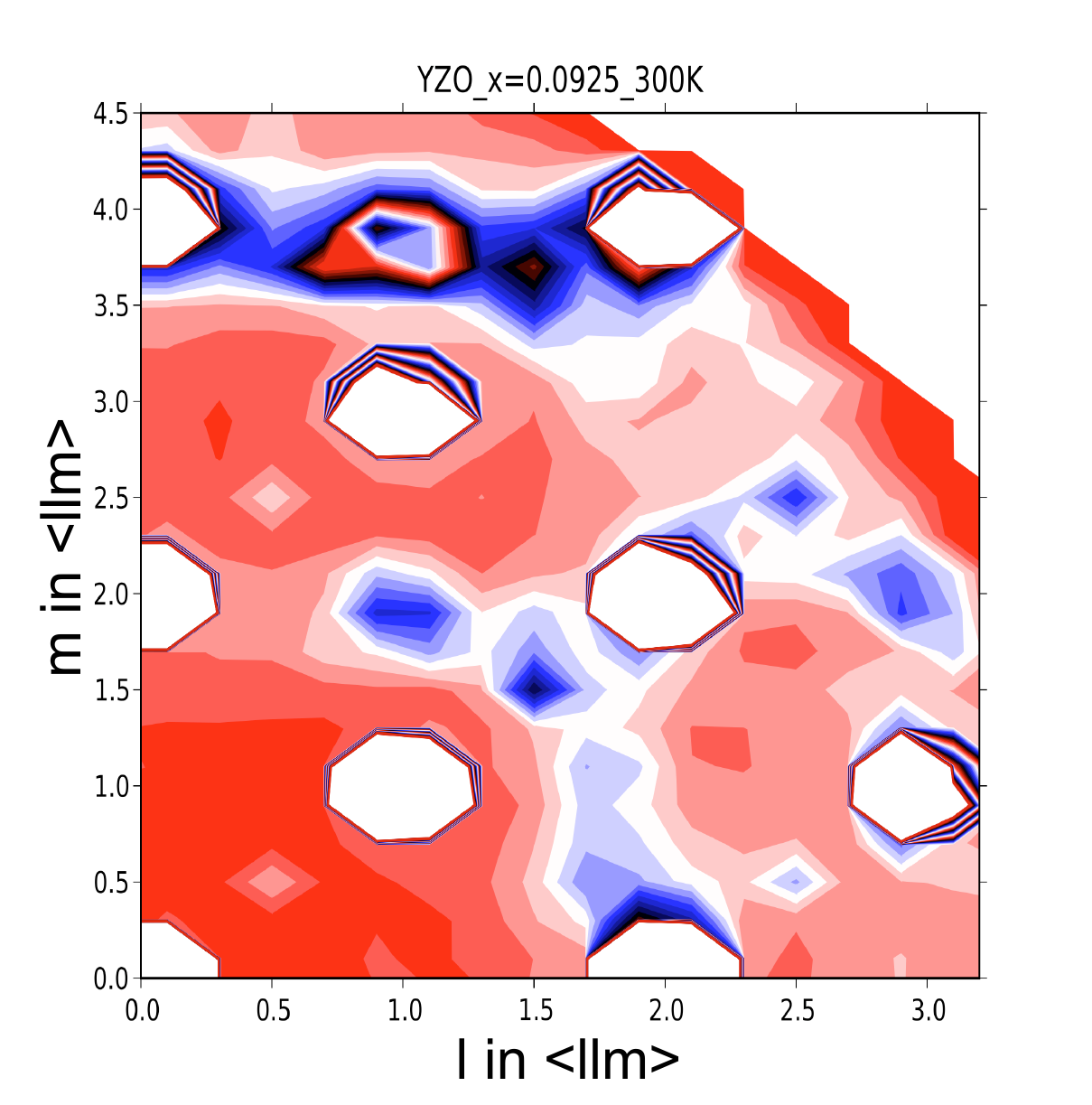}
\includegraphics[width=6.0cm]{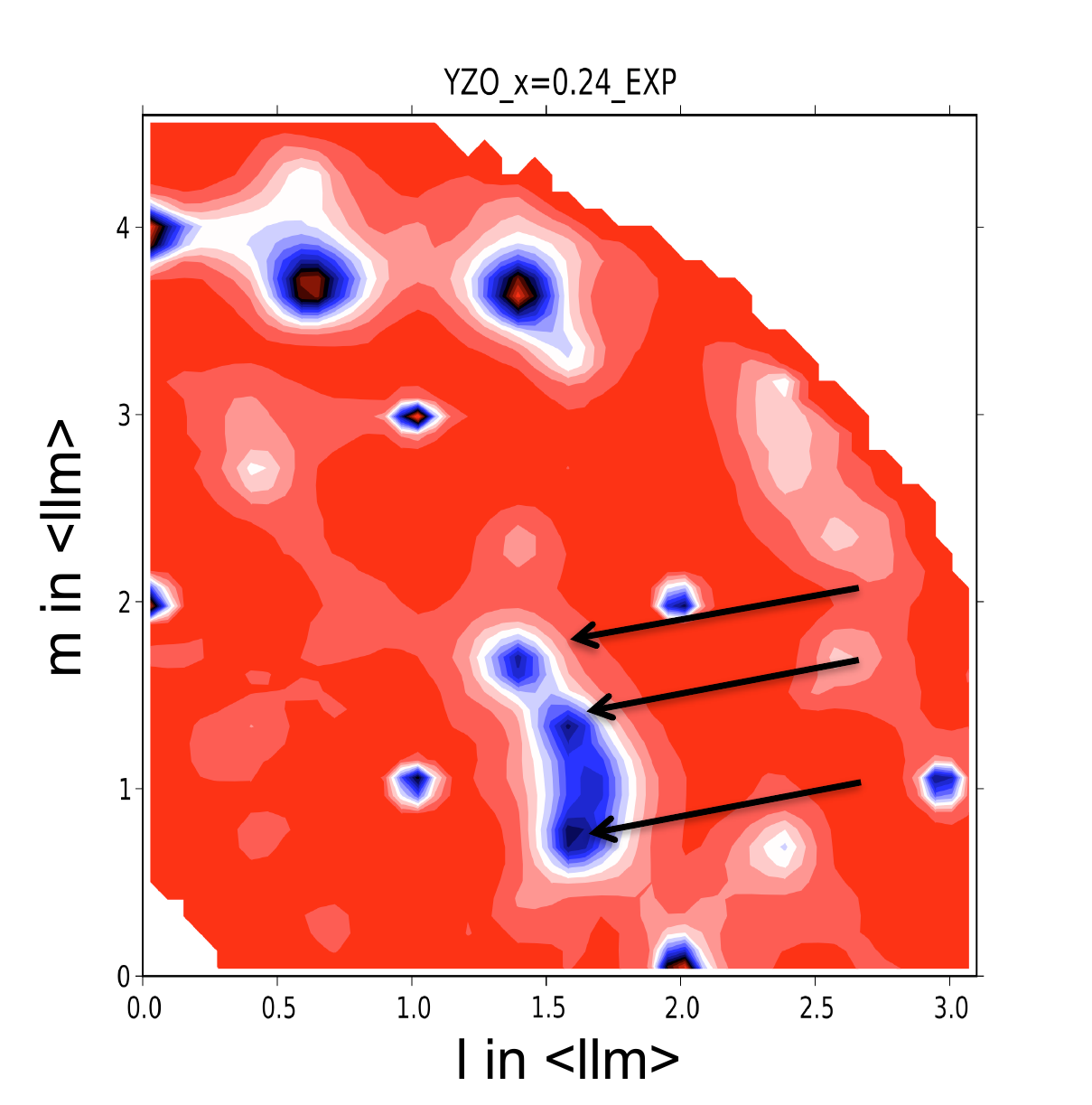}
\includegraphics[width=6.0cm]{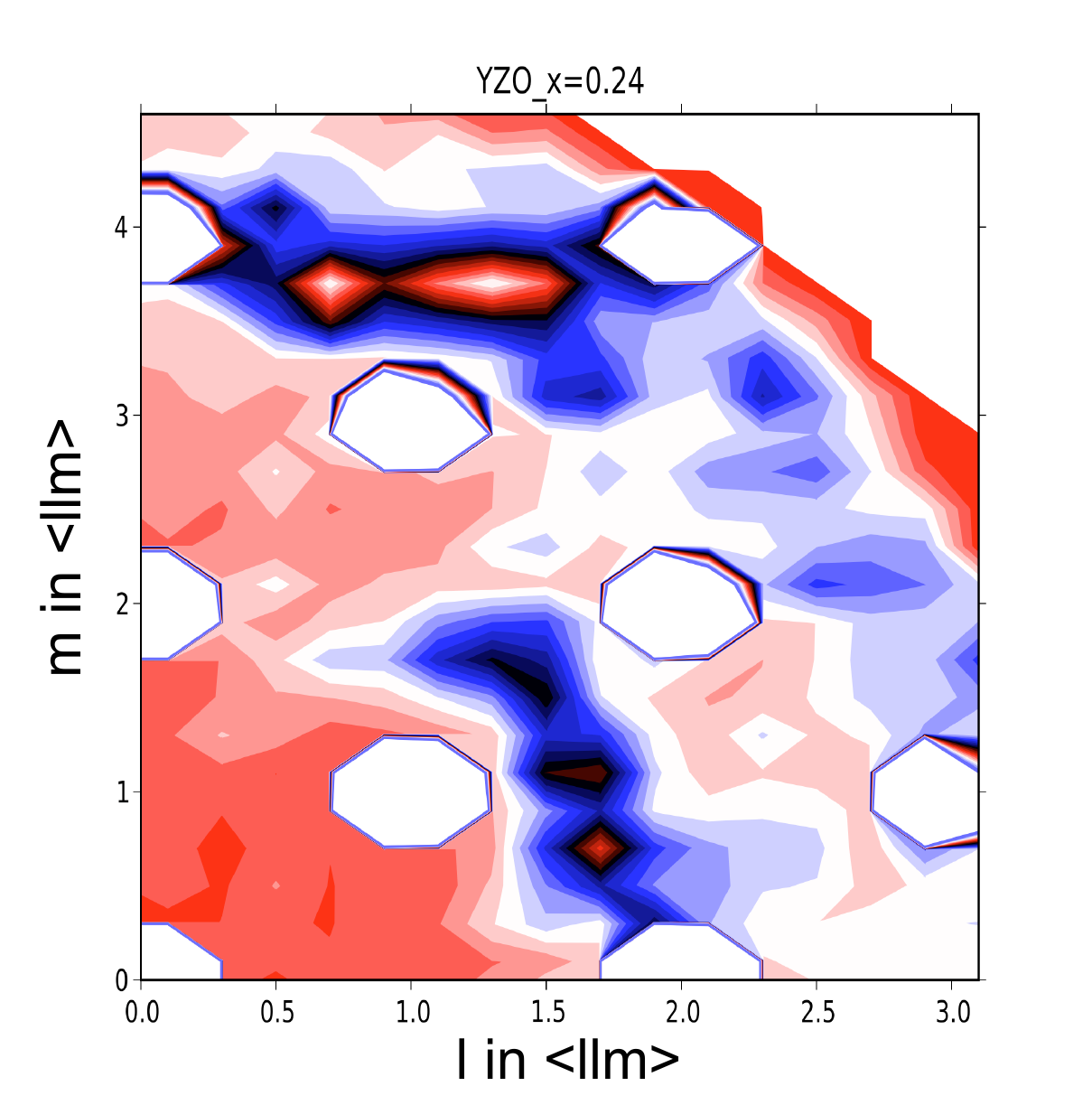}
\end{center}
\caption{\it{\small Experimental (left) and computed (right) neutron diffraction patterns for 9 \% (top) 
 and 24 \% (bottom) YSZ.}}
\label{ND_COMP}
\end{figure}

As we remarked in the Introduction, the most direct evidence for vacancy-vacancy ordering effects in the temperature range of interest for conductivity measurements comes from the single-crystal diffuse neutron scattering data obtained by Goff {\it et al.} \cite{goff1999}. In figure \ref{ND_COMP} we have compared the experimental patterns obtained for  (Y$_2$O$_3$)$_x$ - (ZrO$_2$)$_{1-x}$, with $x$ = 0.09 and $x$ = 0.24 respectively, with patterns calculated in the simulations for very similar compositions. \newline

The $x$ = 0.09 (T = 300 K) composition corresponds to the maximum of the conductivity plot for YSZ. The figure shows a $\langle 1{\bar 1}0 \rangle $ plane in reciprocal space. Sharp intense peaks are seen in the experimental data at the fcc Bragg positions $ \langle 002  \rangle$, $ \langle 004  \rangle$, $ \langle 111 \rangle$, $ \langle 113  \rangle$, $ \langle 220  \rangle$, $ \langle 222 \rangle$, {\it etc.}; in the simulation data the Bragg peaks are centred on the same ${\bf q}$ values but are broader because of the lower resolution. The remainder of the intensity which appears in the pattern is diffuse scattering and is indicative of local structural deformations. At $x$ = 0.09 the most intense diffuse features in the experimental data appear at $ \langle 114  \rangle$ and $ \langle 112  \rangle$, these are forbidden reflections for the fluorite structure. According to the analysis of Goff {\it et a.l} these peaks are a consequence of local tetragonal distortions of the fluorite structure which occur at relatively low concentrations of Y$_2$O$_3$. Recall that in pure ZrO$_2$, a tetragonal phase is more stable than the cubic one and that Y$_2$O$_3$ is added to stabilize the latter. At low Y$_2$O$_3$ concentrations local tetragonal distortions with random orientation occur even within a single crystal. Comparison with the MD data shows that the same diffuse features are present in the simulation data, together with other features which are comparatively weak in the experimental data. This indicates that the tendency to local tetragonal distortion is reproduced in the simulated system: the differences in the intensity distribution and the widths of the diffuse features compared to experiment may be attributed to the small size of our simulation cell which sets a limit to the range of correlation of the tetragonal deformation which can be accommodated and thus leads to broader, weaker diffuse features. Both sets of data also show a broad, weak feature at ${\bf q} = (1.7,1.7,1)$ which is associated with a pattern of lattice deformation about isolated vacancies \cite{goff1999}. \newline

As the Y$_2$O$_3$ concentration is increased to $x$ = 0.24 (T = 800 K), the $\langle 114 \rangle$ and $\langle 112 \rangle$ features disappear and the diffuse scattering becomes dominated by new features (highlighted by the arrows in the bottom part of figure \ref{ND_COMP}) which may be described as belonging to a superlattice at ${\bf G} \pm \langle 0.4,0.4,\pm0.8 \rangle$, where ${\bf G}$ is an fcc Bragg peak position \cite{goff1999}. This pattern has been interpreted as caused by small aggregates of divacancy clusters (vacancies paired along the $\langle 111 \rangle$ direction) packed along the $\langle 112 \rangle$ direction, as found in the Zr$_3$Y$_4$O$_{12}$ compound. These aggregates are typically 15 \AA\ in diameter \cite{goff1999}. This structure is consistent with the most stable arrangement of vacancy pairs found by Bogicevic \cite{bogicevic2003} in his analysis of vacancy-ordering tendencies in YSZ and with our analysis in the preceding paper. Comparison with the simulation data shows that the signature of this particular vacancy-vacancy ordering has been reproduced in the simulation. Interestingly, since our simulations at $x$ = 0.24 were run at T = 800 K, this means that these ordering effects are still present at high temperature and that therefore they will affect the conducting properties of these materials at the temperatures of interest for technological applications. \newline

Although we have only discussed the diffuse scattering results for YSZ itself, similar diffuse scattering patterns were obtained in calculations on ScSZ, for which there is no single crystal data.


 \subsection{Quasielastic scattering}
 \begin{figure}[htbp]
\begin{center}
\includegraphics[width=10cm]{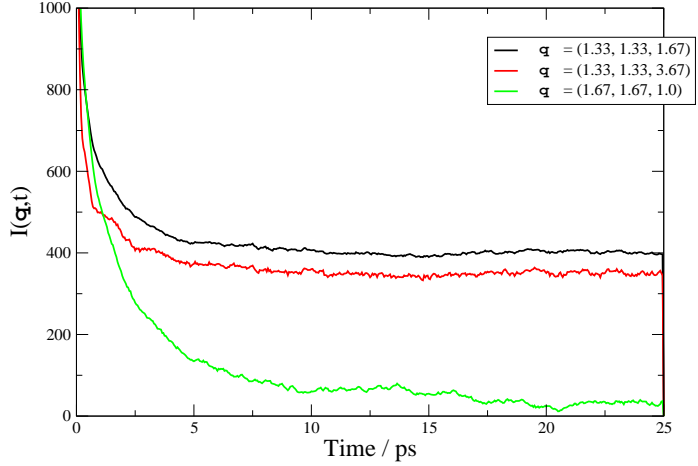}
\end{center}
\caption[Quasielastic scattering in 13\% YSZ]{\it{\small Time scan at constant {\bf q} through the quasielastic diffuse scattering from 13 \% YSZ at 1800 K. }}
\label{Quasielastic}
\end{figure}

As a final comparison with experiment, we now attempt to reproduce the quasielastic data from ref. \cite{goff1999}.   Goff {\it et al.} \cite{goff1999} showed that the relaxation times of the quasielastic scattering at different points in reciprocal space varied significantly. Close to the diffuse scattering feature associated with the isolated vacancies (at ${\bf q}$= (1.67, 1.67, 1) the intermediate scattering function at 1800 K relaxed quite rapidly, on a timescale of 1.3 ps, which gives a timescale for the independent vacancy hopping. However, at ${\bf q}$= (1.33,1.33,1.67) and ${\bf q}$= (1.33,1.33,1.67), where the diffuse scattering is associated with clusters of vacancy pairs, the relaxation was much slower, leading to an unresolved elastic peak in the associated spectra. Goff {\it et al.} deduced from this that those vacancies involved in the clusters moved much more slowly than the free ones, giving compelling evidence for the effect of vacancy-vacancy interactions on diffusion.

The intermediate scattering function at these three ${\bf q}$ values was calculated as explained in the previous section, from a simulation at 1800 K on 13 \% YSZ; these were the conditions of the experimental study. The results are shown in figure \ref{Quasielastic}. It can be seen  that only at  ${\bf q}$ = (1.67, 1.67, 1) does the intermediate scattering function relax to zero on the timescale of the simulation. At  ${\bf q}$ = (1.67, 1.67, 1) the relaxation time is 1.7 ps \symbolfootnote[6]{This was obtained by fitting an exponential decay to the curve in figure \ref{Quasielastic}.}, to be compared with the 1.3 ps seen experimentally. At the other ${\bf q}$ values there would be a significant unresolved "elastic" peak in the spectrum indicating that the defect aggregates responsible for the scattering at these ${\bf q}$ values are immobile on the timescale on which the correlation function has been calculated. The simulations are thus in good quantitative accord with the experimental results.

\section{Analysis of the vacancy ordering}

Although the diffuse scattering comparison discussed above demonstrates that the ordering effects present in the real material are reproduced in the simulations, the {\em degree} of ordering is more conveniently illustrated in real-space, by comparing cation-vacancy and vacancy-vacancy radial distribution function, obtained for different dopant concentrations.  \newline

In the preceding paper, we have already illustrated how the cation-vacancy radial distribution functions provide information about the tendency of vacancies to associate with particular cation types. Our results are consistent with earlier suggestions \cite{kilner1982, khan1998, zacate1999, arachi1999} that the cation-vacancy interactions are driven by the lattice strain associated with the difference in size between the host and dopant cations. The fact that vacancies and trivalent dopants have the same charge, and would therefore be expected to repel, seems to play a much smaller role.  It was shown that in YSZ, where there is a substantial disparity in size between the two cations, there was a high tendency for the vacancies to occur in the first coordination shell of Zr, relative to Y. This can be interpreted as allowing the Zr to reduce its coordination number from the 8 of the normal fluorite structure.  In ScSZ, where the cation sizes are much closer, this tendency was much less marked. The stronger cation-vacancy ordering effect in YSZ provides a natural way of explaining the higher conductivity of ScSZ at the  low dopant level which was the focus of the previous study. However, in view of the weakness of the effect in ScSZ, unless the cation-vacancy becomes much stronger at higher dopant levels, it seems likely that other factors contribute to the drop in conductivity and the appearance of a maximum at relatively low dopant levels. \newline

\begin{figure}[htbp]
\begin{center}
\includegraphics[width=10cm]{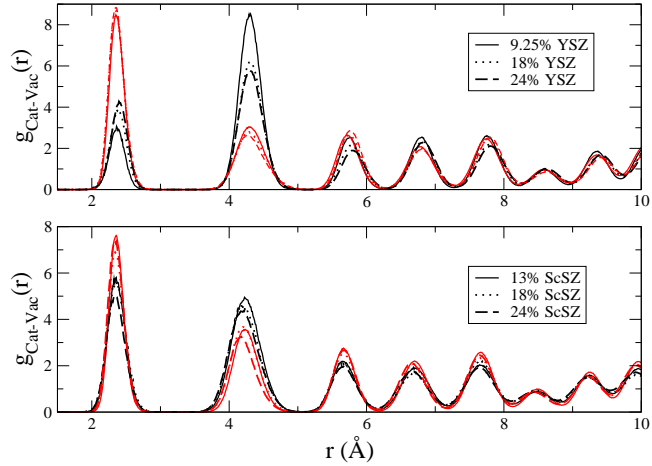}
\end{center}
\caption[Cation-vacancy radial distribution functions for YSZ and ScSZ]{\small \it{Cation-vacancy radial distribution functions for YSZ (top) and ScSZ (bottom) for different dopant concentrations at 1250 K. The red curves are the Zr-Vac rdfs while the black curves represent the dopant cation-vacancy rdfs.}}
\label{Cat-Vac_MSZ}
\end{figure}

In figure \ref{Cat-Vac_MSZ}, we show the cation-vacancy radial distribution functions for YSZ and ScSZ for different dopant concentrations at 1250 K. We choose to show the concentration at which the maximum in the conductivity is observed (9\% and 13\% for YSZ and ScSZ, respectively) and then two higher concentrations (18\% and 24\%). The rdfs for YSZ show that there is a preference for vacancies to be nearest neighbours to the small Zr$^{4+}$ cations, as found in the preceding paper for 11\% YSZ. Interestingly, these binding tendencies become {\em weaker} as more dopant cations are added. This is probably caused by the fact that, as we add more Y$_2$O$_3$, the number of dopant cations increases (as well as the number of vacancies) and it becomes more difficult for the vacancies to avoid the dopant cations. In ScSZ, on the other hand, the vacancies have almost no preference for a certain cation species for all the dopant concentrations, as shown in figure \ref{Cat-Vac_MSZ}, and this preference becomes even weaker as the dopant concentration is increased. 
Figure \ref{Cat-Vac_MSZ} seems therefore to confirm the hypothesis that cation-vacancy interactions alone cannot explain the anomalous behaviour of the conductivity in these materials. These effects are almost absent in ScSZ at the temperatures of interest and they become even less important at higher dopant concentrations. We now focus on the importance of vacancy-vacancy interactions.
\newline

It might be anticipated that the vacancy-vacancy ordering would be affected by the vacancy-cation ordering. For this reason it is convenient to supplement the information on the vacancies in the real systems YSZ and ScSZ with that calculated on the ideal system i-YSZ$_{x}$, where all the cations are identical, so that the only vacancy-ordering effects are caused by the vacancy-vacancy interactions. The vacancy-vacancy rdfs for YSZ, ScSZ  and i-YSZ$_x$ have been studied as a function of the vacancy concentration and temperature. The peaks appear at the positions of the (simple cubic) oxide lattice sites, so that the first peak corresponds to a nearest-neighbour position along $\langle 100 \rangle$, the second to $\langle 110\rangle $, the third to $\langle 111\rangle $ {\it etc.} If the vacancies were randomly distributed over the lattice sites, with no correlations between them, the vacancy-vacancy rdf would have the same appearance as an anion-anion rdf ($g_{a-a}$) for a perfect fluorite system (for instance PbF$_2$) and we have included a plot of $g_{a-a}$ in both figures to illustrate the strength of the vacancy-vacancy correlations which are actually found: the relative intensities of the peaks in $g_{V-V}$, relative to those in $g_{a-a}$, give a guide to the probability of finding vacancies at these separations compared to the random distribution. \newline

\begin{figure}[htbp]
\begin{center}
\includegraphics[width=9cm]{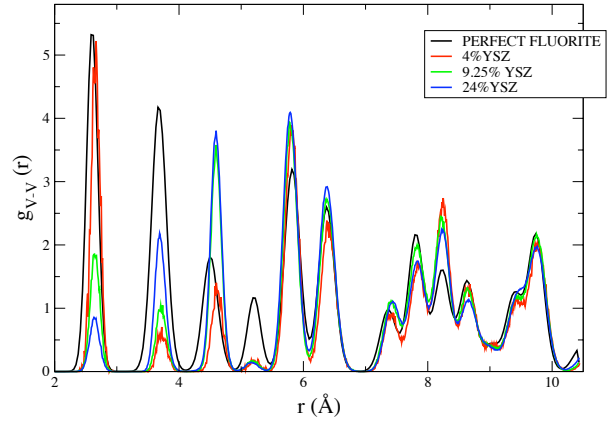}
\includegraphics[width=9cm]{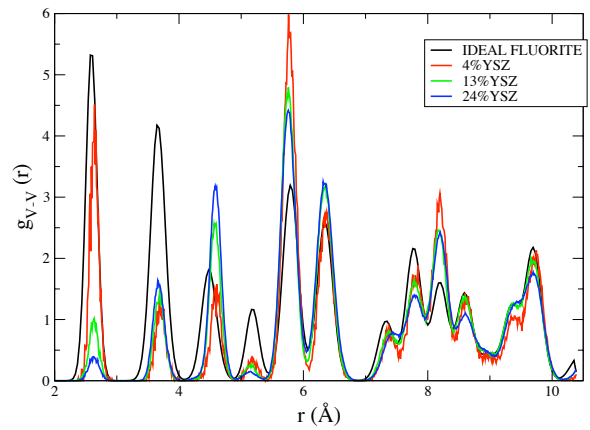}
\includegraphics[width=9cm]{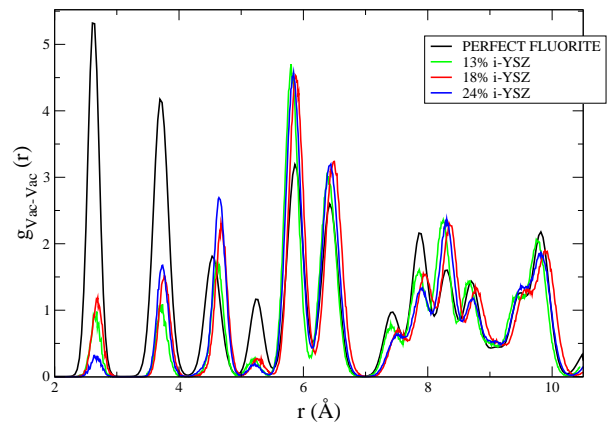}
\end{center}
\caption{\it{\small Vacancy-vacancy radial distribution functions for  (Y$_2$O$_3$)$_x$ - (ZrO$_2$)$_{1-x}$ (top) and  (Sc$_2$O$_3$)$_x$ - (ZrO$_2$)$_{1-x}$ (bottom)  at 1250 K.}}
\label{Vac-Vac_RDF_1250K}
\end{figure}

The data for YSZ are illustrated in the top part of figure \ref{Vac-Vac_RDF_1250K} for a simulation at 1250 K and for three different dopant concentrations, $x$. At small separations it is clear that the $\langle 100 \rangle$ and $\langle 110 \rangle$ relative positions are very unfavourable relative to the random distribution, and that the occupancy of the $\langle 111 \rangle$ position is considerably enhanced. These propensities are consistent with the relative energy order which Bogicevic \cite{bogicevic2003} found in {\it ab initio} studies of the stability of different vacancy pairs at 0 K. Indeed we have already seen in the preceding paper that the enthalpy difference between the <110> and <111> vacancy pairs coincides well with the energy difference found in the ab-initio studies. These propensities are also consistent with the interpretation \cite{goff1999} of the diffuse scattering in the $x$ = 0.24 sample discussed above. Also, these propensities do not seem to change as a function of the dopant concentrations studied here. It is interesting to note that occupancy of a pair of positions separated by the $\langle 111 \rangle$ vector results in a substantial lattice distortion, as seen in the position of the corresponding peak in  $g_{V-V}$ compared to $g_{a-a}$. It is also clear that the vacancy-vacancy correlations extend well beyond the first unit cell and, indeed, the rdf only begins to match that of the random distribution for separations larger than 9 \AA. The $\langle 200 \rangle$ position seems particularly unfavourable, but there is an enhanced occupation of the $\langle 210 \rangle$ and $\langle 211 \rangle$ positions which suggests that the pairs of vacancies are themselves beginning to order. It is well known that at the composition Y$_4$Zr$_3$O$_{12}$, corresponding to $x$ = 0.4, the YSZ system forms a compound which can be described as based on the fluorite structure with ordered vacancy pairs in positions which are themselves ordered along the $\langle 112 \rangle$ direction and this appears to be the tendency which is being picked up in  $g_{V-V}$ even at considerably smaller $x$ values. \newline

The middle and bottom parts of figure \ref{Vac-Vac_RDF_1250K} shows the vacancy-vacancy rdfs for ScSZ and i-YSZ$_x$. These show very similar trends to those observed in YSZ, with vacancies preferring pairing up in the $\langle 111 \rangle$ direction. A comparison between the top and middle part of figure \ref{Vac-Vac_RDF_1250K} shows that these ordering tendencies are quite similar between the two materials and that therefore they are not influenced by the dopant cation species. The implications of this finding are that the vacancy-vacancy ordering tendencies are not affected by the nature of the dopant species and that they are an intrinsic property of the fluorite lattice.
\newline

\section{Discussion and conclusions}
We have established that the strength of the vacancy-vacancy ordering effects is very similar in YSZ, ScSZ and i-YSZ$_x$. Goff {\it et al.} showed that the vacancy clustering affects the conductivity of these materials by reducing the mobility of the vacancies \cite{goff1999} and we have confirmed this idea in the simulated systems. It remains to show at what dopant levels the vacancy-vacancy ordering effects begin to influence the conductivity. \newline

\begin{figure}[htbp]
\begin{center}
\includegraphics[width=10cm]{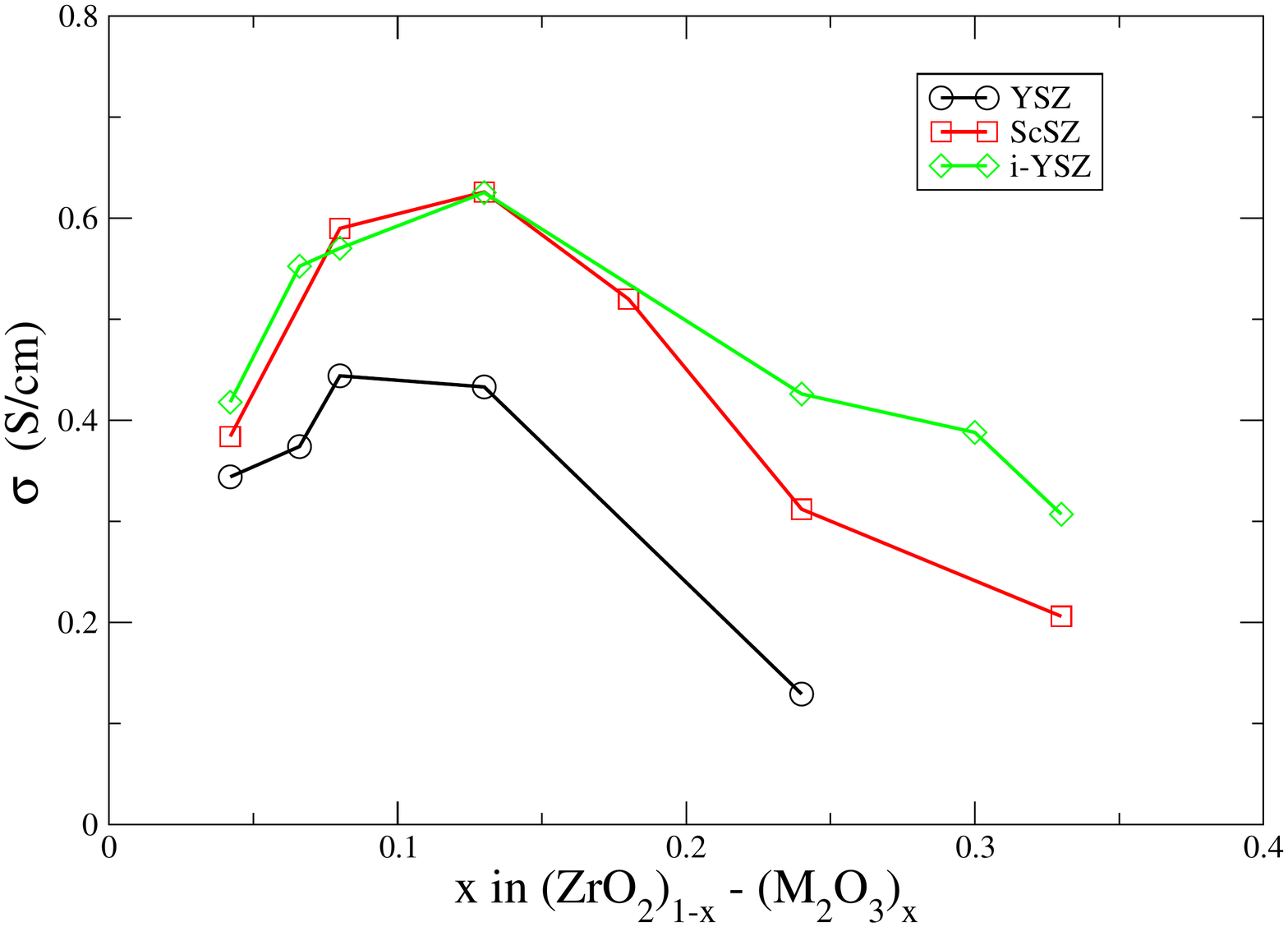}
\end{center}
\caption[Conductivity versus vacancy concentration for i-YSZ$_x$ ]{\small \it {Conductivity versus vacancy concentration for i-YSZ$_x$ , YSZ and ScSZ at 1670 K. The data for i-YSZ$_x$ has been rescaled by a factor of 0.8 to facilitate the comparison. That i-YSZ$_x$ has a slightly higher conductivity, at each $x$, than ScSZ is probably a consequence of the potential which has, on average, a less repulsive short-range interactions (see the discussion in reference \cite{marrocchelli2009b}).}}
\label{IDEAL-CONDUCTOR}
\end{figure}

We can do this conveniently by calculating the $x$-dependence of the conductivity of i-YSZ$_x$, in which there are no dopant-vacancy interactions, and comparing it with that of the real materials. This is shown in figure \ref{IDEAL-CONDUCTOR}. The shape of the conductivity versus $x$ curve for i-YSZ$_x$ follows closely that of ScSZ, especially in the vicinity of the maximum, although the i-YSZ$_x$ values are systematically higher. At higher dopant concentrations, the two curves differ and this may reflect the fact that at these higher concentrations the equalisation of the charges on all the cations in i-YSZ$_x$ exerts a significant effect on the interactions and also the short-range potential in this system is, on average, slightly less repulsive than in the real systems.
\newline


Figure \ref{IDEAL-CONDUCTOR} shows that, even if all the differences between the cation species are removed, a drop in the conductivity is still observed in i-YSZ$_x$ and this happens at 13\%, which is approximately the same vacancy concentration as in the ``real'' materials. This demonstrates that the observed anomalous behaviour is mainly caused by vacancy-vacancy interactions. Vacancy-vacancy interactions are caused by the fact that vacancies are charged and distort the lattice \cite{bogicevic2003}. As more vacancies are introduced in the material, they will try to minimise these interactions by ordering over the fluorite lattice. This limits the mobility of vacancies by correlating their motion, as explained in ref. \cite{pietrucci2008}. Unlike cation-vacancy interactions, which can be reduced by choosing a suitable dopant cation which matches the radius of Zr$^{4+}$, vacancy-vacancy interactions cannot be minimised in such a way. This also implies that this is an {\em intrinsic} effect and therefore a ``property'' of the fluorite lattice. This explains the weak dependence of the maximum's position on the dopant cation's radius observed by Arachi {\it et al.} \cite{arachi1999}. Dopants with larger radii (Y$^{3+}$, Dy$^{3+}$, Gd$^{3+}$) will increase the cation-vacancy interaction and therefore slightly lower the maximum's position, but will not influence very much the vacancy-vacancy interactions, which is the main cause of the anomalous behaviour of the conductivity of these materials. For this reason ScSZ is probably the stabilised zirconia with the highest achievable ionic conductivity, as the Sc$^{3+}$ radius matches that of Zr$^{4+}$ most closely. \newline

\newpage

\section*{Acknowledgements}
DM thanks the Moray Endowment Fund of the University of Edinburgh for the purchase of a workstation. DM also wishes to thank the EPSRC, School of Chemistry, University of Edinburgh, and the STFC CMPC for his PhD funding. STN wishes to thank the EU Research and Technology Development Framework Programme for financial support.

\bibliographystyle{rsc}
\bibliography{rsc}

\ifx\mcitethebibliography\mciteundefinedmacro
  \PackageError{rsc.bst}{mciteplus.sty has not been loaded}
  {This bibstyle requires the use of the mciteplus package.}
\fi
\begin{mcitethebibliography}{35}
\providecommand{\natexlab}[1]{#1}
\mciteSetBstSublistMode{f}
\mciteSetBstMaxWidthForm{subitem}
{(\emph{\alph{mcitesubitemcount}})}
\mciteSetBstSublistLabelBeginEnd
{\mcitemaxwidthsubitemform\space}
{\relax}{\relax}

\bibitem[Hull(2004)]{hull2004}
S.~Hull, \emph{Rev. Prog. Phys.}, 2004, \textbf{67}, 1233--1314\relax
\mciteBstWouldAddEndPuncttrue
\mciteSetBstMidEndSepPunct{\mcitedefaultmidpunct}
{\mcitedefaultendpunct}{\mcitedefaultseppunct}\relax
\EndOfBibitem
\bibitem[Arachi \emph{et~al.}({1999})Arachi, Sakai, Yamamoto, Takeda, and
  Imanishai]{arachi1999}
Y.~Arachi, H.~Sakai, O.~Yamamoto, Y.~Takeda and N.~Imanishai, \emph{{SOLID
  STATE IONICS}}, {1999}, \textbf{{121}}, {133--139}\relax
\mciteBstWouldAddEndPuncttrue
\mciteSetBstMidEndSepPunct{\mcitedefaultmidpunct}
{\mcitedefaultendpunct}{\mcitedefaultseppunct}\relax
\EndOfBibitem
\bibitem[Khan \emph{et~al.}({1998})Khan, Islam, and Bates]{khan1998}
M.~Khan, M.~Islam and D.~Bates, \emph{{JOURNAL OF MATERIALS CHEMISTRY}},
  {1998}, \textbf{{8}}, {2299--2307}\relax
\mciteBstWouldAddEndPuncttrue
\mciteSetBstMidEndSepPunct{\mcitedefaultmidpunct}
{\mcitedefaultendpunct}{\mcitedefaultseppunct}\relax
\EndOfBibitem
\bibitem[Zacate \emph{et~al.}(1999)Zacate, Minervini, Bradfield, Grimes, and
  Sickafus]{zacate1999}
M.~O. Zacate, L.~Minervini, D.~J. Bradfield, R.~W. Grimes and K.~E. Sickafus,
  \emph{Solid State Ionics}, 1999,  243\relax
\mciteBstWouldAddEndPuncttrue
\mciteSetBstMidEndSepPunct{\mcitedefaultmidpunct}
{\mcitedefaultendpunct}{\mcitedefaultseppunct}\relax
\EndOfBibitem
\bibitem[Goff \emph{et~al.}(1999)Goff, Haynes, Hull, Hutchings, and
  Clausen]{goff1999}
J.~Goff, W.~Haynes, S.~Hull, M.~Hutchings and K.~Clausen, \emph{Phys. Rev. B},
  1999, \textbf{59}, 14202\relax
\mciteBstWouldAddEndPuncttrue
\mciteSetBstMidEndSepPunct{\mcitedefaultmidpunct}
{\mcitedefaultendpunct}{\mcitedefaultseppunct}\relax
\EndOfBibitem
\bibitem[Bogicevic and Wolverton(2003)]{bogicevic2003}
A.~Bogicevic and C.~Wolverton, \emph{Phys. Rev. B}, 2003, \textbf{67},
  024106\relax
\mciteBstWouldAddEndPuncttrue
\mciteSetBstMidEndSepPunct{\mcitedefaultmidpunct}
{\mcitedefaultendpunct}{\mcitedefaultseppunct}\relax
\EndOfBibitem
\bibitem[Kilo \emph{et~al.}(2003)Kilo, Argirusis, Borchardt, and
  Jackson]{kilo2003}
M.~Kilo, C.~Argirusis, G.~Borchardt and R.~Jackson, \emph{Phys. Chem. Chem.
  Phys.}, 2003, \textbf{5}, 2219\relax
\mciteBstWouldAddEndPuncttrue
\mciteSetBstMidEndSepPunct{\mcitedefaultmidpunct}
{\mcitedefaultendpunct}{\mcitedefaultseppunct}\relax
\EndOfBibitem
\bibitem[Krishnamurthy \emph{et~al.}(2000)Krishnamurthy, Yoon, Srolovitz, and
  Car]{krishnamurthy2004}
R.~Krishnamurthy, Y.-G. Yoon, D.~J. Srolovitz and R.~Car, \emph{J. Am. Ceram.
  Soc.}, 2000, \textbf{87}, 1821\relax
\mciteBstWouldAddEndPuncttrue
\mciteSetBstMidEndSepPunct{\mcitedefaultmidpunct}
{\mcitedefaultendpunct}{\mcitedefaultseppunct}\relax
\EndOfBibitem
\bibitem[Kushima and Yildiz(2009)]{kushima2009}
A.~Kushima and B.~Yildiz, \emph{ECS Transactions}, 2009, \textbf{25},
  1599\relax
\mciteBstWouldAddEndPuncttrue
\mciteSetBstMidEndSepPunct{\mcitedefaultmidpunct}
{\mcitedefaultendpunct}{\mcitedefaultseppunct}\relax
\EndOfBibitem
\bibitem[Shannon({1976})]{shannon1976}
R.~Shannon, \emph{{ACTA CRYSTALLOGRAPHICA SECTION A}}, {1976}, \textbf{{32}},
  {751--767}\relax
\mciteBstWouldAddEndPuncttrue
\mciteSetBstMidEndSepPunct{\mcitedefaultmidpunct}
{\mcitedefaultendpunct}{\mcitedefaultseppunct}\relax
\EndOfBibitem
\bibitem[Devanathan \emph{et~al.}(2006)Devanathan, Webber, Singhal, and
  Gale]{devanathan2006}
R.~Devanathan, W.~Webber, S.~Singhal and J.~Gale, \emph{Solid State Ionics},
  2006, \textbf{177}, 1251\relax
\mciteBstWouldAddEndPuncttrue
\mciteSetBstMidEndSepPunct{\mcitedefaultmidpunct}
{\mcitedefaultendpunct}{\mcitedefaultseppunct}\relax
\EndOfBibitem
\bibitem[Sawaguchi and Ogawa(2000)]{sawaguchi2000}
N.~Sawaguchi and H.~Ogawa, \emph{Solid State Ionics}, 2000, \textbf{128},
  183\relax
\mciteBstWouldAddEndPuncttrue
\mciteSetBstMidEndSepPunct{\mcitedefaultmidpunct}
{\mcitedefaultendpunct}{\mcitedefaultseppunct}\relax
\EndOfBibitem
\bibitem[Pornprasertsuk \emph{et~al.}(2005)Pornprasertsuk, Ramanarayanan,
  Musgrave, and Prinz]{pornprasertsuk2005}
R.~Pornprasertsuk, P.~Ramanarayanan, C.~Musgrave and F.~Prinz, \emph{J. Appl.
  Phys.}, 2005, \textbf{98}, 103513\relax
\mciteBstWouldAddEndPuncttrue
\mciteSetBstMidEndSepPunct{\mcitedefaultmidpunct}
{\mcitedefaultendpunct}{\mcitedefaultseppunct}\relax
\EndOfBibitem
\bibitem[Martin(2006)]{martin2006}
M.~Martin, \emph{J. Electroceram.}, 2006, \textbf{17}, 765--773\relax
\mciteBstWouldAddEndPuncttrue
\mciteSetBstMidEndSepPunct{\mcitedefaultmidpunct}
{\mcitedefaultendpunct}{\mcitedefaultseppunct}\relax
\EndOfBibitem
\bibitem[Devanathan \emph{et~al.}(2009)Devanathan, Thevuthasan, and
  Gale]{devanathan2009}
R.~Devanathan, S.~Thevuthasan and J.~Gale, \emph{Phys. Chem. Chem. Phys.},
  2009, \textbf{11}, 5506\relax
\mciteBstWouldAddEndPuncttrue
\mciteSetBstMidEndSepPunct{\mcitedefaultmidpunct}
{\mcitedefaultendpunct}{\mcitedefaultseppunct}\relax
\EndOfBibitem
\bibitem[Sato \emph{et~al.}(2009)Sato, Suzuki, Yashiro, Kawada, Yugami,
  Hashida, Atkinson, and Mizusaki]{sato2009}
K.~Sato, K.~Suzuki, K.~Yashiro, T.~Kawada, H.~Yugami, T.~Hashida, A.~Atkinson
  and J.~Mizusaki, \emph{Solid State Ionics}, 2009, \textbf{180}, 1220\relax
\mciteBstWouldAddEndPuncttrue
\mciteSetBstMidEndSepPunct{\mcitedefaultmidpunct}
{\mcitedefaultendpunct}{\mcitedefaultseppunct}\relax
\EndOfBibitem
\bibitem[Norberg \emph{et~al.}(2010?)Norberg, Hull, Peng, Irvine, Marrocchelli,
  and Madden]{norberg2010}
S.~Norberg, S.~Hull, L.~Peng, J.~Irvine, D.~Marrocchelli and P.~Madden,
  \emph{submitted to Journal of Material Chemistry}, 2010?,  405403\relax
\mciteBstWouldAddEndPuncttrue
\mciteSetBstMidEndSepPunct{\mcitedefaultmidpunct}
{\mcitedefaultendpunct}{\mcitedefaultseppunct}\relax
\EndOfBibitem
\bibitem[Irvine \emph{et~al.}({2000})Irvine, Feighery, Fagg, and
  Garcia-Martin]{irvine2000}
J.~Irvine, A.~Feighery, D.~Fagg and S.~Garcia-Martin, \emph{Solid State
  Ionics}, {2000}, \textbf{{136}}, {879--885}\relax
\mciteBstWouldAddEndPuncttrue
\mciteSetBstMidEndSepPunct{\mcitedefaultmidpunct}
{\mcitedefaultendpunct}{\mcitedefaultseppunct}\relax
\EndOfBibitem
\bibitem[Garcia-Martin \emph{et~al.}({2000})Garcia-Martin, Alario-Franco, Fagg,
  Feighery, and Irvine]{garcia2000}
S.~Garcia-Martin, M.~Alario-Franco, D.~Fagg, A.~Feighery and J.~Irvine,
  \emph{{CHEMISTRY OF MATERIALS}}, {2000}, \textbf{{12}}, {1729--1737}\relax
\mciteBstWouldAddEndPuncttrue
\mciteSetBstMidEndSepPunct{\mcitedefaultmidpunct}
{\mcitedefaultendpunct}{\mcitedefaultseppunct}\relax
\EndOfBibitem
\bibitem[Garcia-Martin \emph{et~al.}({2005})Garcia-Martin, Alario-Franco, Fagg,
  and Irvine]{garcia2005}
S.~Garcia-Martin, M.~Alario-Franco, D.~Fagg and J.~Irvine, \emph{{Journal of
  Material Chemistry}}, {2005}, \textbf{{15}}, {1903--1907}\relax
\mciteBstWouldAddEndPuncttrue
\mciteSetBstMidEndSepPunct{\mcitedefaultmidpunct}
{\mcitedefaultendpunct}{\mcitedefaultseppunct}\relax
\EndOfBibitem
\bibitem[Pietrucci \emph{et~al.}(2008)Pietrucci, Bernasconi, Laio, and
  Parrinello]{pietrucci2008}
F.~Pietrucci, M.~Bernasconi, A.~Laio and M.~Parrinello, \emph{Phys. Rev. B},
  2008, \textbf{78}, 094301\relax
\mciteBstWouldAddEndPuncttrue
\mciteSetBstMidEndSepPunct{\mcitedefaultmidpunct}
{\mcitedefaultendpunct}{\mcitedefaultseppunct}\relax
\EndOfBibitem
\bibitem[Kilner and Brook(1982)]{kilner1982}
J.~Kilner and R.~Brook, \emph{Solid State Ionics}, 1982, \textbf{6}, 237 --
  252\relax
\mciteBstWouldAddEndPuncttrue
\mciteSetBstMidEndSepPunct{\mcitedefaultmidpunct}
{\mcitedefaultendpunct}{\mcitedefaultseppunct}\relax
\EndOfBibitem
\bibitem[Norberg \emph{et~al.}(2009)Norberg, Ahmed, Hull, Marrocchelli, and
  Madden]{norberg2009a}
S.~Norberg, I.~Ahmed, S.~Hull, D.~Marrocchelli and P.~Madden, \emph{J. Phys.:
  Condens. Matter}, 2009,  215401\relax
\mciteBstWouldAddEndPuncttrue
\mciteSetBstMidEndSepPunct{\mcitedefaultmidpunct}
{\mcitedefaultendpunct}{\mcitedefaultseppunct}\relax
\EndOfBibitem
\bibitem[Marrocchelli \emph{et~al.}({2009})Marrocchelli, Madden, Norberg, and
  Hull]{marrocchelli2009a}
D.~Marrocchelli, P.~A. Madden, S.~T. Norberg and S.~Hull, {SOLID-STATE
  IONICS-2008}, {506 KEYSTONE DRIVE, WARRENDALE, PA 15088-7563 USA}, {2009},
  pp. {71--78}\relax
\mciteBstWouldAddEndPuncttrue
\mciteSetBstMidEndSepPunct{\mcitedefaultmidpunct}
{\mcitedefaultendpunct}{\mcitedefaultseppunct}\relax
\EndOfBibitem
\bibitem[Marrocchelli \emph{et~al.}(2009)Marrocchelli, Madden, Norberg, and
  Hull]{marrocchelli2009b}
D.~Marrocchelli, P.~Madden, S.~Norberg and S.~Hull, \emph{J. Phys.: Condens.
  Matter}, 2009, \textbf{21}, 405403\relax
\mciteBstWouldAddEndPuncttrue
\mciteSetBstMidEndSepPunct{\mcitedefaultmidpunct}
{\mcitedefaultendpunct}{\mcitedefaultseppunct}\relax
\EndOfBibitem
\bibitem[Madden \emph{et~al.}(2006)Madden, Heaton, Aguado, and
  Jahn]{madden2006a}
P.~Madden, R.~Heaton, A.~Aguado and S.~Jahn, \emph{J. Mol. Struct.: THEOCHEM},
  2006, \textbf{771}, 9--18\relax
\mciteBstWouldAddEndPuncttrue
\mciteSetBstMidEndSepPunct{\mcitedefaultmidpunct}
{\mcitedefaultendpunct}{\mcitedefaultseppunct}\relax
\EndOfBibitem
\bibitem[Wilson \emph{et~al.}({2004})Wilson, Jahn, and Madden]{wilson2004}
M.~Wilson, S.~Jahn and P.~Madden, \emph{{JOURNAL OF PHYSICS-CONDENSED MATTER}},
  {2004}, \textbf{{16}}, {S2795--S2810}\relax
\mciteBstWouldAddEndPuncttrue
\mciteSetBstMidEndSepPunct{\mcitedefaultmidpunct}
{\mcitedefaultendpunct}{\mcitedefaultseppunct}\relax
\EndOfBibitem
\bibitem[Jahn and Madden(2007)]{jahn2007b}
S.~Jahn and P.~Madden, \emph{Phys. Earth Planet. Inter.}, 2007, \textbf{162},
  129--139\relax
\mciteBstWouldAddEndPuncttrue
\mciteSetBstMidEndSepPunct{\mcitedefaultmidpunct}
{\mcitedefaultendpunct}{\mcitedefaultseppunct}\relax
\EndOfBibitem
\bibitem[Marrocchelli \emph{et~al.}({2009})Marrocchelli, Salanne, Madden,
  Simon, and Turq]{marrocchelli2009c}
D.~Marrocchelli, M.~Salanne, P.~A. Madden, C.~Simon and P.~Turq,
  \emph{{MOLECULAR PHYSICS}}, {2009}, \textbf{{107}}, {443--452}\relax
\mciteBstWouldAddEndPuncttrue
\mciteSetBstMidEndSepPunct{\mcitedefaultmidpunct}
{\mcitedefaultendpunct}{\mcitedefaultseppunct}\relax
\EndOfBibitem
\bibitem[Marrocchelli \emph{et~al.}(2010)Marrocchelli, Salanne, and
  Madden]{marrocchelli2010a}
D.~Marrocchelli, M.~Salanne and P.~A. Madden, \emph{Journal of Physics:
  Condensed Matter}, 2010, \textbf{22}, 152102\relax
\mciteBstWouldAddEndPuncttrue
\mciteSetBstMidEndSepPunct{\mcitedefaultmidpunct}
{\mcitedefaultendpunct}{\mcitedefaultseppunct}\relax
\EndOfBibitem
\bibitem[Martyna \emph{et~al.}(1994)Martyna, Tobias, and Klein]{martyna1994a}
G.~Martyna, D.~Tobias and M.~Klein, \emph{J. Chem. Phys.}, 1994, \textbf{101},
  4177--4189\relax
\mciteBstWouldAddEndPuncttrue
\mciteSetBstMidEndSepPunct{\mcitedefaultmidpunct}
{\mcitedefaultendpunct}{\mcitedefaultseppunct}\relax
\EndOfBibitem
\bibitem[Castiglione \emph{et~al.}({2001})Castiglione, Wilson, Madden, and
  Grey]{castiglione2001}
M.~Castiglione, M.~Wilson, P.~Madden and C.~Grey, \emph{{JOURNAL OF
  PHYSICS-CONDENSED MATTER}}, {2001}, \textbf{{13}}, {51--66}\relax
\mciteBstWouldAddEndPuncttrue
\mciteSetBstMidEndSepPunct{\mcitedefaultmidpunct}
{\mcitedefaultendpunct}{\mcitedefaultseppunct}\relax
\EndOfBibitem
\bibitem[Subbarao and Ramakrishnan(1979)]{subbarao1979}
E.~Subbarao and T.~Ramakrishnan, \emph{Fast Ion Transport in Solids}, New York:
  Elsevier/North Holland, 1979, pp. 653--656\relax
\mciteBstWouldAddEndPuncttrue
\mciteSetBstMidEndSepPunct{\mcitedefaultmidpunct}
{\mcitedefaultendpunct}{\mcitedefaultseppunct}\relax
\EndOfBibitem
\bibitem[Nakamura and Wagner(1986)]{nakamura1986}
A.~Nakamura and J.~J. Wagner, \emph{J. Electrochem. Soc.}, 1986, \textbf{1542},
  133\relax
\mciteBstWouldAddEndPuncttrue
\mciteSetBstMidEndSepPunct{\mcitedefaultmidpunct}
{\mcitedefaultendpunct}{\mcitedefaultseppunct}\relax
\EndOfBibitem
\end{mcitethebibliography}

\end{document}